%% file: 0_main.tex
\documentclass[sn-apa]{sn-jnl}

\jyear{2022}
\usepackage{comment}
\usepackage{amsmath,amssymb}
\usepackage{graphicx}
\usepackage{array}
\usepackage{titlesec}
\usepackage{lipsum}
\usepackage{url}
\usepackage[justification=centering]{caption}
\usepackage{wrapfig,lipsum,booktabs}

\titleformat{\paragraph}
{\normalfont\normalsize\bfseries}{\theparagraph}{1em}{}
\titlespacing*{\paragraph}
{0pt}{3.25ex plus 1ex minus .2ex}{1.5ex plus .2ex}
\newcommand\blfootnote[1]{%
  \begingroup
  \renewcommand\thefootnote{}\footnote{#1}%
  \addtocounter{footnote}{-1}%
  \endgroup
}

\begin{document}

\title[Intelligent Energy Management Systems: A Review]{Intelligent Energy Management Systems: A Review}

\author*[1]{
\fnm{Stavros} 
\sur{Mischos}}
\email{mischossg@csd.auth.gr}

\author[2]{
\fnm{Eleanna} 
\sur{Dalagdi}}
\email{edalagdi@csd.auth.gr}

\author[1]{
\fnm{Dimitris}
\sur{Vrakas}}
\email{dvrakas@csd.auth.gr}

\affil*[1]{\orgdiv{School of Informatics}, 
\orgname{Aristotle University of Thessaloniki}, \orgaddress{\city{Thessaloniki}, \country{Greece}}}

\abstract{
Climate change has become a major problem for humanity in the last two decades. One of the reasons that caused it, is our daily energy waste. People consume electricity in order to use home/work appliances and devices and also reach certain levels of comfort while working or being at home. However, even though the environmental impact of this behavior is not immediately observed, it leads to increased CO2 emissions coming from energy generation from power plants. Confronting such a problem efficiently will affect both  the environment and our society. Monitoring energy consumption in real-time, changing energy wastage behavior of occupants and using  automations with incorporated energy savings scenarios, are  ways to decrease  global energy footprint. In this review, we study intelligent systems for energy management  in residential, commercial and educational buildings, classifying them in two major categories depending on whether they provide direct or indirect control. The article also discusses what the strengths and weaknesses are, which optimization techniques do they use and finally, provide insights about how these systems can be improved in the future.
}


\maketitle

\input{1_intro}
\input{2_motivation}
\input{3_related}

\input{4_efficiency}

\input{5_iems}
\input{6_discussion}
\input{7_conclusion}

\newpage

\bibliography{sn-bibliography}

\end{document}

%% file: 1_intro.tex
\section{Introduction}\label{sec:intro}

\noindent
\blfootnote{\textbf{Acknowledgement:} This research has been co-financed by the European Regional Development Fund of the European Union and Greek national funds through the Operational Program Competitiveness, Entrepreneurship and Innovation, under the call RESEARCH–CREATE–INNOVATE (project code:95699 - Energy Controlling Voice Enabled Intelligent Smart Home Ecosystem).}

\par Nowadays, electrical energy plays a vital role in various aspects of our life. However, the lack of ecological awareness along with the absence of energy-friendly infrastructures has led into increased energy consumption and  waste. According to estimates of the United States Energy Information Administration \cite{1}, 40\% of the annual CO2 emissions are directly related to the electricity consumption. Out of these emissions, 40\% of them concern residential and commercial consumers, and as \cite{2} and  \cite{3} mentioned, it is possible to achieve 20\% savings if we use power more efficiently. Therefore,  electricity consumption  and wastage redution can offer environmental and financial benefits to our society.

 \par Different approaches and systems have been proposed in the literature that aim  to reverse  climate change and global warming. Intelligent energy management systems with incorporated automations is a promising approach towards the solution of these environmental problems. These systems convert a conventional home or building into a "smart" version of it. Smart Homes and Buildings, according to \cite{6}, include automations systems which provide the ability to monitor and control various services such as, lighting and heating-ventilation-airconditioning (HVAC), or devices such as fridges, ovens and washing machines. The set of installed sensors, actuators and smart devices constitute an Internet-of-Things (IoT) subsystem. When users are surrounded by microcontrollers and smart devices, they  follow the paradigm of \textit{Ubiquitous or Pervasive Computing}. When Artificial Intelligence (AI) methodologies enable the interaction of people with these devices, the environment embodies \textit{Ambient Intelligence (AmI)} (\cite{28}).  
 
\par These environments play an important role in the \textit{Smart Grid}. Smart grids consist of two parts, the \textit{supply-side} and the \textit{demand-side}, which optimize the energy production, transmission, distribution and consumption (\cite{4}). Smart homes are a necessity for the demand-side of these grids because even if the supply-side is successfully optimized, a faulty demand-side, e.g. a conventional home/building, will decrease the total effectiveness of the  system.

\par An immediate  conversion of all residential and commercial buildings  from conventional to smart, is a costly and time-consuming procedure. Even if governments around the world wanted to carry out this plan, the high deployment costs remain an impediment (\cite{8}, \cite{9}). Therefore, research was expanded towards lower or no-cost energy saving solutions based on information and communcation technologies (ICT) (\cite{10}).

\par Research and developement of energy management systems focused on new technologies that embody energy savings, and materials that decrease wastage of energy. However, the same attention wasn't given at users' behavioral change.  \cite{3}, suggested that in order to improve the awareness on energy waste habits, consumers must firstly  monitor their power consumption and then manage it after receiving appropriate advice. Governments and Non-Governmental Organizations support and facilitate energy-efficiency changes. However the impact of simple saving tips and peer devices' comparison is low because of the wrong time and place these occur (\cite{11}). End-users must alter  their routine completely and adopt an environmental friendly behavior. (\cite{13})

\par Recommendation systems are information systems that assist users to discover personalized content, based on their preferences (\cite{14}). They are used in  many different real-world scenarios (\cite{15}) and recently some implementations  were applied  into energy profile reshaping (\cite{16}, \cite{17}) using deep learning algorithms (\cite{18}), data mining techniques (\cite{12}), behavioral analytics and human decision-making processes to develop context-aware systems (\cite{20}, \cite{19}). Despite the fact that these systems emerged in the mid 90's (\cite{21}, \cite{22}), \cite{23} mentioned that, the field of energy saving recommendation systems is still unexplored.

\par This work provides a comprehensive review of Intelligent Energy Management Systems (IEMS) in the literature over the last decade. Our research was focused on surveying two classes of systems. The first one includes intelligent systems with automated controls, aiming at energy wastage reduction and the second one systems with human actuators, aiming at behavioral modifications. From our point of view, these systems are two sides of the same coin, therefore an overall review of the state of the art was necessary. Furthermore, it is important to understand where they differ and also in which environment is each class best suited. To accomplish that, it is necessary to break down IEMS in their components and analyze their distinguished features.

\par The remaining  of our study is organized as follows. Section \ref{sec:motivation} presents our motivation towards studying and comparing these two types of energy management systems. Section \ref{sec:related} discusses related work on this topic and refers to surveys performed on smart environments and recommendation systems for energy efficiency. Section \ref{sec:efficiency} provides background information of the energy efficiency topic from a researchers' perspective. In section \ref{sec:iems}, we present necessary specifications for an intelligent energy management system and an overview of their architectures. Subsequently, section \ref{sec:discussion} presents a discussion and an analysis of the advantages of each class, their problematic issues and some suggestions for future research. Finally, section \ref{sec:conclusion} concludes our findings.

%% file: 2_motivation.tex
\section{Motivation}
\label{sec:motivation}

\par Energy management systems comprise a promising solution towards energy wastage reduction. The variety of studies on smart environments, and the plurality of   algorithms and techniques developed over the last decade for automations and recommendations' optimizations, are proofs of how important these systems are in our effort to reverse climate change and global warming. During our research, we noticed that in current  literature, every discussion about smart environments involved mostly systems with integrated automations. Nevertheless, new systems emerged recently which incorporate recommendations mechanisms, aiming at end-users' awareness improvement rather than in automations. Therefore, we believe that a survey is needed in the literature merging automations and recommendations as Intelligent Energy Management Systems, discussing  advantages and disadvantages between these two approaches.

\par Firstly, we want to present an overview of the major influential factors which cause energy wastage and also refer to some general approaches towards energy efficiency. In this way, the reader can understand why both the automations and recommendations approaches must be considered as appropriate technologies to reduce society's energy footprint.

\par Secondly, we want to present the architecutre of IEMS, therefore is is necessary to define their specific functional and non-functional requirements. Moreover, we will refer to state-of-the-art approaches of each component and also establish a novel classification for these systems classifying them as Direct and Indirect Control IEMS. Our goal is to provide the reader with understanding on why   these systems are designed that way and what the differences  between direct and indirect control are.

\par Finally, our intention is to draw some conclusions about the advantages of each class, their major problematic issues, and also some future research orientations in this field.

%% file: 3_related.tex
\section{Related Work}
\label{sec:related}

\par In this section, we present surveys and reviews that are related to IEMS (Table \ref{tab1}). Our intention is to provide the reader an extensive look in the field of Energy Management Systems in Residential, Educational and Commercial Buildings. One could read the work of \cite{26} in order to understand which are the general approaches to energy efficiency and also the main architectural, technological and algorithmic aspects of an energy management system. \cite{5} proposed a similar architecture that also incorporates smart appliances, while \cite{51} presented a more abstract  architecture for IoT-based systems, consisting of three layers: a \textit{Perception layer}, a \textit{Network layer} and an \textit{Application} layer. Finally, \cite{54} dealt with a review of the state-of-the-art Building Energy Management Systems(BEMS) focusing on three model  approaches: White box, Black box and Grey box models. They also performed a comparative analysis of the factors that have the highest impact in energy consumption.

\par When IEMS are  installed, they provide AmI at the environment. According to  \cite{53}, there are various definitions about AmI depending on the features that are included. These environments offer environmental, comfort, safety and financial benefits. AmI is also an umbrella term which  applies into technologies embedded into a physical space to create an invisible user interface augmented with AI (\cite{24}). They presented an comprehensive survey on Ambient Intelligence (AmI) and Ambient Assistive Living (AAL) while referring to the state-of-the-art AI techniques and methodologies to implement these systems. 

\par  An interesting study was performed by  \cite{55} reviewing load scheduling controllers  which integrate AI techniques such as,  artificial neural networks (ANNs), fuzzy logic, adaptive neural fuzzy inference and heuristic optimization. \cite{5} discussed about the demand-side management, i.e. the collection of techniques applied to reduce energy costs on the consumption-side and improve energy efficiency. Furthermore, they  discussed about dominant scheduling methodologies which are grouped into five categories. Additional energy saving techniques are presented by  \cite{4}, including statistical models, cloud computing-based solutions, fog computing, smart-metering-based architectures and also some IoT inspired solutions.  \cite{68} reviewed methods employed to model various aspects of residential energy management systems. Moreover, they discussed about complexity in such systems and   presented also an overview of techniques for scheduling approaches, as well as a classification in mathematical programming, meta-heuristic search and heuristic scheduling techniques. Finally,  a recent study by \cite{63} summarized research opportunities created by open issues in the field such as, blockchain-enabled IoT platforms for distributed energy management, deep learning models to handle, use and evaluate big energy data,  peer-to-peer energy trading and demand-side energy management, context-aware pervasive future computing, resilience-oriented energy management, forecasting models, user comfort and real-time feedback systems as well as, Internet of Energy (IoE)-based energy management.

\par Multi-agent systems (MAS) is an approach used to model components of IEMS. An interesting article was published by \cite{56} reviewing state-of-the-art developments in MAS and how they are used to solve energy optimization problems. They  discussed about the types of MAS architectures and also the reasons why they must be used as modeling tools. \cite{64} proposed a multi-agent architecture of distributed intelligence to solve the complex and dynamic decision process of optimal energy management. Their architecture  was based on four groups of agents: control and monitoring, information, application and management and optimization agents. According to the definition of MAS (\cite{58}) an agent is comprised by a coupling of perception, reasoning and acting components.  \cite{57} conducted a comprehensive literature review  aiming to identify and characterize the reasoning systems in MAS-based smart environments and also presented the strengths and limitations of them.

\par Recommendation systems are an important research field since the mid 90's. Their goal is to help users find online content based on personalized preferences using collaborative or content-based filtering along with AI techniques such as, association rules, clustering, decision trees, k-nearest neighbor, neural networks, regression, etc.(\cite{59}). Over the last decade, the research community began to integrate recommendation modules into the components of smart environment systems to persuade users to adopt a more eco-friendly behavior. An extensive literature survey was conducted by \cite{23} on energy saving recommendation systems in buildings, discussing how they evolved and also providing a taxonomy based on the nature of the recommender engine, their objectives, computing platforms and evaluation metrics. Furthermore, a critical analysis was also conducted to understand what the limitations of these systems are when they aim in energy efficiency. Additionally, \cite{60} studied a set of recommendations published by companies and agencies, and designed micro-models to estimate how popular recommendations affect energy savings and conducted also a followup study to understand which types of recommendations were chosen and adopted more often.

\par From a theoretical point of view,  \cite{11} analyzed  the barriers towards the adoption of technologies for efficient energy usage and introduced ways to overcome them. Moreover, \cite{66} indicated in their meta-review, that automation and control technologies are only "one piece of the puzzle". The other one  is  systems that embed humans in the procedure of excessive consumption reduction using behavioral change techniques through recommendations and feedback. They suggested that the integration between techno-centric and user-centric approaches in more holistic implementations will be more effective.


\begin{table}[ht]
\begin{center}
\begin{minipage}{\textwidth}
\begin{tabular}{@{}p{4cm}lp{7cm}@{}}
\toprule
Ref. Number & Year  & Description \\
\midrule
\cite{4}   & 2021  & State-of-the-art solutions for energy management  \\

\cite{23}   & 2021   & Analysis of recommendation systems for energy  efficiency in buildings \\

\cite{24} & 2021   &  Overview of the field of Ambient Intelligence  \\

\cite{63} & 2021 &   Compilation of latest developments and research orientations on intelligent energy management  \\

\cite{5} & 2020   &  State of the art review of energy management systems in residential buildings   \\

\cite{66} & 2020   & Discussion between techno-centric and user-centric approaches on energy management systems    \\

\cite{11} & 2019  &   Discussion on barriers to the adoption of technologies for energy efficiency  \\

\cite{17} & 2019   &   Users' micro-moments as influential factors of energy consumption behaviors  \\

\cite{13} & 2018  &   Importance of altering users' behavior  \\

\cite{54} & 2018   &    Review on building energy management systems focusing on three model types   \\

\cite{55} & 2018   &  Review of demand response techniques, smart technologies and scheduling controllers   \\

\cite{56} & 2018   &  Multi-agent systems as modelling tools for energy optimization applications   \\

\cite{51} & 2017   & Overview on various aspects of Internet of Things systems  \\

\cite{68} & 2015   &  Analysis of methods for modeling home energy management systems   \\

\cite{26} & 2014   &  Presentation of architectural, technological and algorithmic aspects of intelligent energy management systems   \\

\cite{64} & 2013   &   Proposal of multi-agent architectures for energy management  \\

\cite{59} & 2012   &   Review on trending topics of recommendation systems  \\

\botrule
\end{tabular}
\caption{Previous Works Related to Intelligent Energy Management Systems}\label{tab1}%
\end{minipage}
\end{center}
\end{table}

%% file: 4_efficiency.tex
\section{Efficient Energy Consumption}
\label{sec:efficiency}

\subsection{Influential Factors}

\par The first step to achieve energy waste reduction is to understand where it originates from. According to \cite{12}, there are four major influential factors of this phenomenon: 

\begin{itemize}
    \item \textbf{Building Characteristics:} Construction materials and insulation levels are  obvious factors that increase energy waste in all types of buildings. \cite{69} conducted a research on performance gaps in energy consumption, revealing that recent buildings, constructed with modern materials, consume less energy than recently renovated older buildings. Furthermore, a difference between actual and theoretical (simulated) consumption was also noticed.
    
    \item \textbf{Occupants Behavior:} Occupants affect the overall energy consumption, especially in residential buildings (\cite{70}). Even in buildings with  the same energy labeling, discrepancies can occur in consumption, depending on heater/cooler set temperature, hot water wastage, requirements of indoor environmental quality, lighting usage, etc.

    \item \textbf{System Efficiency and Operation:} Many buildings or households are equipped either with low  efficiency or outdated appliances and devices. Systems' efficiency affects dramatically the total power consumption, as well as, neglected appliances such as oven, microwave, washing and drying machines.

    \item \textbf{Climatic Conditions:} Outdoor temperature, solar radiation, humidity and wind velocity can affect energy consumption, especially combined with the aforementioned factors. Even though, these conditions cannot be managed, it is important to realize how they increase  energy wastage in order to search for effective solutions.

\end{itemize}

\subsection{Approaches towards energy efficiency}

\par In order to confront  climate change, society has to adopt a more energy efficient mindset.  \cite{71} identified four approaches towards energy efficiency: \textit{user awareness about energy consumption}, \textit{reduction of standby consumptions}, \textit{plan and scheduling of flexible activities} and \textit{adaptive control}. 

\par Firstly, user awareness is a straightforward way to achieve energy wastage reduction (\cite{26}). Providing appropriate feedback, advice and recommendations along with detailed information about daily power consumption and total cost can encourage users to follow a more eco-friendly behavior (\cite{3}). However, aggregated measurements of energy consumptions make it difficult to understand which device or behavior causes the biggest waste (\cite{26}). Moreover, various studies (\cite{23}, \cite{60}, \cite{72}) show that behavioral change is still an ineffective strategy in the long term, therefore more research is required.

\par Secondly, standby devices and appliances are hidden  sources of energy wastage. It was first identified as a new challenge in the early '90s (\cite{73}) when analysts began to study the number of appliances that were "leaking" electricity.  According to  \cite{74}, any plugged-in device in standby mode can consume some amount of energy, but the increased number of such devices in households and buildings lead to substantial increments in the total consumption. TVs, PCs, Coffee Machines and Printers, are some of the devices in every home, consuming energy without being used for long periods of time (\cite{26}). An interesting study was conducted by \cite{75}, showing that education was positively correlated with the reduction of standby energy consumption. Moreover, households with children were using less multi-sockets and were less likely to waste standby energy, whereas high-income households were correlated with higher energy consumption.

\par Thirdly, activity planning and scheduling using modern smart automations systems can offer reduction of energy consumption during energy demand peak (\cite{26}). Scheduling activities, offers financial benefits when energy fares vary between day and night. Furthermore, as \cite{76} proposed,  users can also save energy when a house is connected, through a hybrid energy system, into battery storage units.

\par Finally, another approach to reduce wasted energy is the installation of adaptive control mechanisms. HVAC and lightning systems waste energy in order to preserve user's comfort. However, the incorporation of  AI techniques such as user-presence detection, behavior prediction or reinforcement learning control can tune activation times of these services to avoid unnecessary consumption (\cite{26}, \cite{77}).

%% file: 5_iems.tex
\section{Intelligent Energy Management Systems (IEMS)}
\label{sec:iems}

\par A number of computer-aided tools and technologies were proposed in the last decade in order to effectively optimize energy consumption in our daily life.  According to  \cite{26}, each system or model that was developed must fulfill some basic functional and nonfunctional requirements. Each system has to perceive the environmental conditions of the place it will be installed, use the input data to learn users' habits, behaviors, preferences, consumptions per device or appliance and also detect or predict existing context. Moreover, it must provide a way for the users to monitor the consumption and at the same time interact with them using notifications to gather feedback and commands. Finally, it should have the ability to modify its environment through actuation after planning optimized sequences of actions that will both reduce energy wastage and satisfy comfort preferences.
\par In respect to the non-functional requirements, these systems have to maintain intrusiveness at low level in the matter of interaction with the user and the physical infrastructure. Furthermore, scalability and extensibility is desirable in such cases, meaning that the level of abstraction during the design should be high. Also, an intelligent system have to be easily deployed by the users and not not require installation from an expert. In the software engineering part of the  implementation, the principle of modularity is really important to avoid problematic behaviors of the system. Finally, it is required to be interoperable, with respect to physical devices and other software components. 

\par
\begin{figure}[ht]
    \centering
    \includegraphics[scale = 0.3]{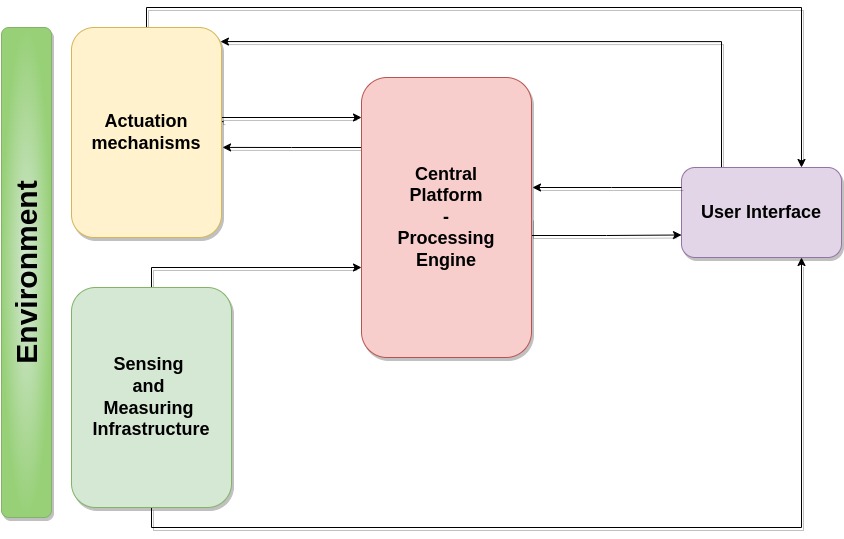}
    \caption{Intelligent Energy Management System}
    \label{fig:IEMS}
\end{figure}

 Energy management systems are developed in a unique way fulfilling the aforementioned requirements following the approaches of the previous section and also following a specific framework architecture (\cite{5}, \cite{26}). The main components of an IEMS are depicted in Fig. \ref{fig:IEMS}:

\begin{enumerate}
    \item Sensing and Measuring Infrastructure 
    \item Actuation mechanisms
    \item Processing Engine
    \item User Interfaces
    
\end{enumerate}

\par Information that emerges from the sensing components is saved and processed by the Processing Engine. The  engine is the specialized subsystem with components responsible for the process of  all acquired data  and also performs the optimization tasks based on the end-users  preferences. It should also learn and recognize occupants' activity patterns, communicate with the actuators and manage anomalies or outlier events. When decisions are made, they are transferred into the actuators to modify each appliance or device contextually. Along with the action commands, sometimes, the process engine  provides recommendations to the end-users through the user interface to change behaviors that affect the total energy consumption. All these modifications are focused on the persuasion of a smaller energy footprint of our society, however economic impacts remain also a motive (\cite{31}). Furthermore, through the user interface, users have 
access to graphs showing daily consumptions. The most common form of a user interface is a computer or smartphone application.

\par A widespread approach to model state-of-the-art energy management systems is Multi-Agent Systems (MAS). MAS architecture is often used as a tool to model  subsystems of an IEMS and is composed of multiple interacting intelligent agents (\cite{78}). Each agent can be considered as "Intelligent" because it incorporated AI techniques such as decision-making or machine learning algorithms.

\par According to  \cite{79}, each agent in a multi-agent system has some important characteristics:
\begin{itemize}
    \item \textbf{Autonomy}: The ability to be at least partially independent, self-aware and autonomous.
    \item \textbf{Local view}: The perception of the agent has boundaries and no agent  has a global view. Otherwise, the agent will not be able to process that large amount of information.
    \item \textbf{Decentralization}: No control authority exists inside the MAS.
\end{itemize}

\par  \cite{79} called the ability of an agent to act at a local level, "Sphere of Influence". Each agent in a MAS has the ability to interact within a specific range. However, there are "spheres" which coincide, rising depency relationships and creating a unified model. Because of that, MAS architectures are considered appropriate to model IEMS.  \cite{56} argued that MAS are commonly used as models because of the communication, coordination and cooperation capabilities of the agents, and also because this design provides robustness to the system, when different tasks are assigned to each agent.

\subsection{Sensors and Measuring Infrastructure}

\par Sensors and measurement devices are installed on every smart environment, providing  data about temperature, humidity and luminance levels, whereas different sensors are monitoring the presence of occupants. There are two types of IoT devices used for these tasks: \textit{Custom-made }and \textit{Commercial}. Arduino or Raspberry Pi microcontrollers are used by researchers to create custom modules that fit specific requirements, but in large-scale applications, commercial ones are a preferable option because of the default unified communication protocols.

\subsubsection{Custom-made Sensors}

\par The most common sensors on these applications are the power consumption meters.  \cite{30} used SEN-11005 components on a microcontroller NodeMCU  to build a  custom energy monitoring device. \cite{97}  built from scratch their own circuit for an electronic sensor consisting of several sub-circuits for, voltage and current metering, voltage regulation and  operational amplification. This sensor was incorporated within an Arduino UNO that was processing the input data which were transfered through an ESP8266 chip to an application. Furthermore, \cite{98} created  an open-source power monitoring system, designed around the Digital Universal Energy Logger (DUEL) Node.  \cite{102} also designed their own smart plug using Zigbee protocol. Finally,  \cite{101} used ACS712 and ZMPT101B for current and voltage measurements, respectively.

\par Temperature and humidity sensors were also commonly used in energy management systems. \cite{96},  \cite{31} and \cite{30} used DHT-22 sensors to receive real-time contextual information from the environment. DHT-22 sensor can measure both temperature and humidity levels. On the contrary,  \cite{99} used an LM35 temperature-only sensor. LM35 has an analog communication protocol, while DHT-22 has one-wire. Therefore, LM35 is faster in data transmission  but it is more sensitive to noise. Also, \cite{100} used LM35 in an ambient intelligent system with an Intel Galileo board.

\par  Light and motion sensors are  extremely important components for systems that aim to reduce energy wastage. Rooms and spaces that  remain unoccupied tend to have increased consumptions due to switched on lights. \cite{18} and  \cite{30} used a TSL2561 Adafruit sensor for light monitoring and HC-SR501 for motion sensing.  \cite{96} also included photoresistor sensors and motion sensors with passive infrared (PIR) in their custom sensor-board and  \cite{99}, used a light dependent resistor (LDR) that reduces its resistance when light hits the surface of it.

\par  \cite{103} created EnAPlug, a multi-sensor smart plug with the ability to switch on/off devices, and monitor power, reactive power, voltage and current. It also included four sensors for temperature, humidity, outside temperature and a door opener detector. 

\subsubsection{Commercial Sensors}

\par Many companies develop smart plugs that are used in energy management systems  for environmental sensing, containing usually multiple sensors on a single device. These devices are utilized in smart homes and buildings and are easier to be used by the average user. Furthermore, researchers are selecting these devices when performing large scale experiments to save time from building custom sensor boards.

\par In their work,  \cite{32} used the following smart plugs with metering abilities and on/off control: the DSP-W125 by D-Link, the SP-2101W by Edimax and the TP-link HS110. The DSP plug has also the ability to monitor temperature.  \cite{34} employed Fibaro 4-in-1 sensor at the site of the experiment to receive data about temperature, humidity, luminosity, motion and presence simultaneously. \cite{44} used digitalSTROM systems which were acting as power meters communicating with other nodes inside a smart environment.

\par \cite{33} utilized three multi-sensors Aeon Gen5 to detect movement, read temperature, luminance and relative humidity values. Furthermore, they used a smart plug  Aeon Smart Switch 6 to control devices and measure instant consumed power and energy and also provide with power consumption graphs for appliances. Last, another Aeon Gen6 multi-sensor was installed that could also provide ultraviolet light sensing data and also home energy meter was installed in the fuse box to measure instant consumption and energy for the entire home without noise.

\subsection{Actuators}

\subsubsection{Direct Control}

\par Actuators are the components of an IEMS that execute decisions and commands in order to perform actions so as to optimize power consumption. There are two possible ways to interact with the devices and appliances. The first one is a set of electronic actuators. Actuators are electrical components that interact with the appliances following the decisions of the process engine after the optimization is performed. All systems with the ability to modify their environment using actuators are called \textit{\textbf{Direct Control IEMS}}. This term  encloses every system able to process data, take decisions and execute them on its own, without the intervention of an human being. 

\par \textit{Elettra} was an innovative system proposed by \cite{36, 25}, allowing users to monitor their power consumption. It incorporated AmI techniques and algorithms to successfully measure and forecast consumptions of devices providing also direct control to sockets using smart plugs and sensors.  \cite{37}  proposed the framework Smart IHU, an application with two components, a Manager and a Rule app. Their actuation infrastructure implementation had custom sensor boards and Z-Wave devices providing automations based on preferable statistics selected by the users. 
\par  \cite{118} implemented an energy management system in a smart meter device. They incorporated a fuzzy logic controller to perform automated actions on the appliances which were divided into two groups, a group of low power devices such as, consumer electronics and multimedia equipment, and a group of medium and high power devices such as, HVAC, water heaters and washing machines.
\par Another platform with direct controls was implemented by  \cite{115}. Their system was designed to minimize the costs per day of a home by optimally scheduling operations. To achieve this, it included controllable household appliances (smart devices) such as pool pump, dish washer, washing machine, clothes dryer, coffee machine, dehumidifier and bread maker. The users were selecting preferred time range for operations and were providing information about their lifestyle.

\subsubsection{Indirect Control}
\par Some  recent approaches put humans in the position of the actuator forming human-in-the-loop  architectures. Their purpose is to  change the behavior of the end-users to stop energy wasting habits. In order to accomplish that, they use recommendations  engines to send suggestions and advice through interfaces in order to motivate people to act optimally.  These systems can also be extended to propose replacements of inefficient devices and appliances that waste energy (\cite{5}). They are called \textit{\textbf{Indirect Control IEMS.}}

\par There are different types of recommendations. The most typical of them are the \textit{personal resources} \textit{recommendations} which advise occupants to turn the lights or the  HVAC appliances off in empty rooms or shut down idle devices such as,  computers and printers. However, three more types of recommendations have been proposed by \cite{18}. \textit{Move recommendations} encourage occupants to change their working/living space to reduce services' requirements. \textit{Schedule change recommendations} are extensions of move recommendations aiming to shift the period of time an occupant spends within a space and not the duration. Finally,  \textit{coerce recommendations} suggest to the  building managers when to force people to evacuate rooms if the occupancy is small in relation to the size of the space. 

\par   \cite{38, 31,39} designed (EM)$^3$, a framework aiming at occupants' behavioral change. Using recommendations from its engine, REHAB-C, the human actuator is getting trained by repetition to behave efficiently. ReViCEE by \cite{7} follows the same logic by predicting energy consumption ratings and offering personalized recommendations, stimulating user-engagement towards energy conservation and sustainability. 

\par \cite{33} designed a modular platform named SHE (Smart Home Environment). The on-premises control was executed using some Z-Wave controllers, TKB Wall dimmer to control dimmable lights and  the Aeon Z-Sticks, Gen-5 and S2, each designed to function on a specific location (US or Europe) depending on the allowed radio frequency for such devices. SHE was advising inhabitants about how they can improve their lifestyle and reduce the costs of energy consumption. Therefore, it provided notifications on a mobile device to motivate them to remotely turn on and off services. 

\par \cite{65} used the framework CAFCLA  to develop a recommendation system for usage in homes to promote efficient energy usage. The system was identifying behavioral patterns and along with CAFCLA's real-time localization system and wireless sensor network it was used to provide personalized recommendations(\cite{8}).
\par KNOTES was a system developed by  \cite{114} that was proposing to the users how to change their life style using  notifications, in order to save energy. The system was taking their personal data such as consumption, owned appliances, percentage of advice acceptance and evaluation history to find appropriate suggestions.
\par Gamified approaches seemed also promising in terms of indirect control, especially in education facilities, due to the increased user engagement they provide, through an achievement system with rewards and leaderboards \cite{34, 113}.

\subsection{Processing Engine}

\par Processing engine of an IEMS is designed to optimize the energy usage on each compartment of a smart environment and manage the actions that have to be performed. After years of research on this field, different techniques have been developed. The majority of the state-of-the-art systems employ Rule Engines,  Data and Pattern mining algorithms, Machine Learning and Deep Learning models 

\subsubsection{Rule Engine} 

\par The most frequently encountered technique on IEMSs is  \textit{Rule Engines}. \cite{35} introduced a general-purpose Rule Engine that pushes notifications or reports to the end-users based on a resource graph model. In (EM)$^3$ (\cite{39}) a goal-based context-aware rule-based system (REHAB-C) was implemented with a rule mining algorithm, a process responsible to gather data about frequency of users actions.  \cite{34}, developed an event-driven rule process on a gamified system aiming to reduce energy-wasting behaviors  where each challenge assigned to the end-users is represented by a specific rule.  \cite{40, 37}, implemented a Rule App in their application in the form of a Hybrid Intelligent Agent. The agent had two interchangeable layers, the deliberative and the reactive. The reactive layer applies and maintains all energy-saving policies, while the deliberative layer incorporates a reasoner, based on defeasible logic  to manage conflicting rules, responsible to optimize energy consumption while maintaining users' comfort. A similar rule-based architecture using defeasible logic was also implemented by \cite{41}. \cite{118}  designed a system that combined a rule-based implementation along with a fuzzy logic algorithm, incorporated on a smart meter to perform direct control. Last,  \cite{34} proposed an event-based rule engine to change energy waste behaviors in public buildings.

\subsubsection{Machine Learning and Data Mining}

\par Smart environments produce  a lot of  data by the IoT components. Even though this data is processed in order to control the environment remotely, until recently they were rarely used from the system to train the models and achieve autonomy.  \cite{43}, designed a  system that takes advantage of contextual  meta-data that originate from  smart devices, using extra-trees classifiers, a technique that combines machine learning and data mining. That way, the dimensionality of the produced data was reduced, without loss of important features. Subsequently, an artificial neural network was trained to complete a context-aware engine, with a continuous learning capability based on feedback from the end-users. 

\par Data mining techniques are also used to monitor inhabitants' usage patterns.  \cite{44}, proposed a sequential pattern mining algorithm aiming on smart environments that predict future needs of their inhabitants. Thus, the system could avoid actions that lead to a comfort decrease. 

\par Another system for energy management based on  mining algorithms was developed by \cite{45}. Their system provided personalized recommendations about turning on and off appliances at specific timestamps based on household profiles produced by association rule mining apporaches, such as Apriori and FP-Growth and also sequential pattern mining apporaches like RuleGrowth, TRuleGrowth, CMRules, ERMiner and CMDeo from a library created by 
rob\cite{46}.

\subsubsection{Deep Learning}

\par On a smart environment where automated actions must be performed, human activities must be monitored. Recognizing patterns of a room occupant will provide necessary information to the back-end system, leading in more effective predictions. Deep learning  techniques such as, convolutional and recurrent neural networks, showed great performance compared to others on human activity recognition \cite{27,29}.  All this information combined with specific sensor measurements can also grant a context model. Using these models, anomalies and outliers can be detected which would affect energy consumption. Moreover, on more sophisticated systems, using all above can lead to residents' identification. That way the system can initialize optimizations and actions based on resident's profile. The aforementioned actions require  complicated calculations, therefore  Deep Neural Networks were employed (\cite{33}).

\par  The problem of the efficient energy consumption consists of two sub-problems,  the non-intrusive load monitoring (NILM) and the energy load forecasting (ELF), which were resolved using deep learning models (\cite{33}). NILM is a method used to monitor the energy profile of an environment and extract information about appliances consumption by disaggregating the total power consumption (\cite{42}). On the other hand, ELF is used to forecast patterns on energy consumption and detect anomalies that might  increase energy consumption (\cite{33}).

\par Another deep learning method used for energy saving is the deep reinforcement learning (DFL).  \cite{18} used a DFL agent, trained along with the end-users' decisions. For each successful reduction of consumption, the agent  received a reward  aimed at maximizing the amount of energy saved. \cite{116} developed such a model, based on Markov Decision Processes (MDP) to control the temperature of domestic hot water. Their goal was to reduce the consumption by optimizing the usage of energy produced by photo-voltaic panels. \cite{117} suggested also a model using MDP to schedule optimally HVAC appliances and the energy storage system of a smart home. Finally,  \cite{52}, proposed a DFL model that incorporated human feedback in the objective function and human activity data in the reinforcement learning part of it to enhance  optimization of energy.

\subsection{User Interface}

\par IEM systems include necessarily a User Interface (UI) to allow interaction between them and the users. First of all, UI displays information about total power consumption or consumption per appliance. Secondly, it provides a mean for indirect or direct control of the devices in a smart space. Moreover, it is the only way for the users to change comfort parameters in direct control systems, schedule functions and set rules. Finally, the interface platform sends notifications stimulating recommended behaviors and receive feedback.

\par Nowadays, interfaces used by smart systems range from simple command-line environments, SMS texts  to  smartphone and smartwatch applications. There are also differences on the approaches of a user interface, meaning that, it could be a simple one just for system manipulation, or a complicated gamified environment, especially in systems aimed at behavioral changes.

\subsubsection{Monitoring and Management Applications}

\par A standard characteristic on every user interface application is the monitoring component. It  usually consists of statistical graphs about consumptions or expenses. \cite{47} designed a visualized performance graph of a building allowed the users to compare measurements from two different time periods. Moreover, the interface was used to provide alerts if outliers were detected. A simple monitoring agent  was also introduced by \cite{48} providing statistical data about hourly consumptions  and the ability to compare different days' consumptions while it  displayed estimated costs for the total  kWh consumed and calculated CO2 emissions amount for a month.

\par A different approach in monitoring was given by \cite{37}, where every room had its own section at the implemented Rule App, displaying statistics such as current power consumption, temperature, humidity, luminance and CO2 levels, and an indication about motion detection. However, their implementation allowed direct manipulation of these variables so that the agent could adjust  comfort levels.  \cite{40}, a new agent with a GUI was introduced allowing the management of the rule engine through a user-friendly interface. A similar approach was implemented also by \cite{36,49} with three main menu choices, allowing manual activation of physical devices, setting of rules, actions and scenarios and also measurement of energy consumption and displaying along with real-time and stored data.

\par  \cite{38} implemented  (EM)$^3$, a system with an end-user web application contained on-line daily consumption monitoring, displaying also indoor and outdoor levels of temperature and humidity, providing at the same time recommendations about energy efficiency. Additionally it had also a control menu with switches to modify devices' activities. A similar user interface was prototyped also by \cite{45}.

\subsubsection{Smart devices notifications}
 
Many IEMS are using smartphones and smartwatches as their user interface. Mobile devices offer the convenience of monitoring and managing consumption on the go.  \cite{31} and \cite{44}, used  the default text message services in order to push  notifications with recommendations expecting feedback from the user to proceed in further actions.

\par \cite{50} proposed a real time energy usage tracking software, along with a web app, two mobile applications for iOS and Android devices and also one for Android wearables.  The main dashboard of the system was in a more compressed form in the smartphone app, whereas the smartwatch app displayed only energy footprint breakdown and notifications when alarms were triggered.

\subsubsection{Gamified Approaches}

\par For systems that aim at behavioral changes, gamified interfaces and applications seem to have impact.  \cite{34, 113}  proposed a gamified approach was introduced based on challenges for public buildings. End-users had a mobile application  plus an NFC chip installed on their smartphones  that showed available challenges, e.g. use staircases instead of the elevator or turn off your PC before leaving your office, and rewarded points in case of completion. The user interface contained also a leaderboard, where users were competing each other.

\begin{sidewaystable}
\sidewaystablefn%
\begin{center}
\begin{minipage}{\textheight}
\caption{Complete state-of-the-art Intelligent Energy Management Systems Implementations}\label{tab2}
\begin{tabular*}{\textheight}{@{\extracolsep{\fill}}lcccp{7cm}@{\extracolsep{\fill}}}
\toprule%
Authors & System's Name  & Control Type & Sensors & Processing Engine   \\
\midrule
\vspace{0.3cm}

 \cite{30} & (EM)$^3$  & Indirect  & Custom & Rule-based + Recommendations Techniques (REHAB-C)     \\ \vspace{0.3cm}


 \cite{118} & --- & Direct  & Custom & Rule-based using Fuzzy Logic  \\\vspace{0.3cm}

 \cite{36} & Elettra  & Direct  & Commercial  & Rule-based using Defeasible logic \\\vspace{0.3cm}

 \cite{8} & --- &  Indirect  & Commercial & Neural Networks + Recommendations Techniques(CAFCLA framework)\\\vspace{0.3cm}

 \cite{7} & ReViCEE   & Indirect  & Custom  & Recommendations Techniques  \\\vspace{0.3cm}

 \cite{113} & ChArGED  & Indirect  & Commercial & Rule Engine \\\vspace{0.3cm}

 \cite{33} & SHE   & Indirect  &  Commercial  &  Various Deep Learning Models \\\vspace{0.3cm}

 \cite{37} & Smart IHU & Direct  & Commercial  & Rule-based using Defeasible logic \\\vspace{0.3cm}

 \cite{50} & ePrints   & Indirect  & Custom  & Deep Reinforcement Learning (recEnergy)   \\\vspace{0.1cm}\\

\botrule
\end{tabular*}
\end{minipage}
\end{center}
\end{sidewaystable}

%% file: 6_discussion.tex
\section{Discussion}
\label{sec:discussion}

\par In our survey, we elaborated over the topic of Intelligent Energy Management Systems. Our goal in this work was to report state-of-the-art approaches of the area (Table \ref{tab2}). However, our work differs from previous surveys because it merges smart automation and recommendation systems  under the umbrella term of IEM systems, considering them as two individual sub-classes, whereas in current literature,  most articles about Home/Building Energy Management Systems (HEMS/BEMS) refer to systems with incorporated automations.

\par In this article, first, we referred to the major influential factors that increase energy consumption and wastage, and also presented some general approaches that should be followed in order to pursue energy efficiency. Second, we provided an IEMS architecture overview, based on some functional and non-functional requirements which arise from the aforementioned influential factors and discussed about the state of the art of these systems. Third, we showed that each IEMS can be classified as a Direct or Indirect Control IEMS based on the type of actuation it incorporates. When a system contains automation mechanics and the end-users choose only environmental preferences, the system is controlling the environment directly. On the contrary, when the end-users receive recommendations to perform actions, the performed control is indirect. More recent studies proved that recommendations systems provide improvements in terms of energy savings. For this reason, we classified automations and recommendation systems as Direct and Indirect Control IEMS, respectively, a novel classification that addresses the lack of a unified structure and helps new researchers to obtain a more accurate overview of the field. Following, we will discuss about the advantages of each IEMS control type (Table \ref{tab3}), as well as their major open issues, providing also some future research orientations.

\subsection{Advantages of Direct Control IEMS} 

\par Direct control IEM systems have the ability to automate procedures and actions. Their major advantage is that the occupants of a smart environment are not obliged to alter their routine. Once preferences are set, the system receives  input data, optimizes and sends action commands to all required appliances, without the need for further interaction. For example, when an occupant leaves a room and no human presence is detected, the lights can be turned off. Similarly, the TV can automatically be turned off if the room is empty.

\par Another argument in favour of direct control is the plan and scheduling ability, allowing users to set long-term ecological or economical  goals on the system. Moreover, scheduling techniques can offer consumption shifting on various operations to find optimal time frames in order to reduce energy demand and cost.

\par Furthermore, smart environments with  automations can protect occupants from emergency situations that can occur, e.g. an excessive power consumption caused by electricity "leakage" from a faulty appliance. A leaky device is  wasting energy and is also hazardous for the  occupants because it can cause accidents, such as fire or electrocution. Automated actuators have the ability to turn off appliances when necessary, reducing the energy wastage and offering also a safer environment.

\par Finally, fully automated IEMS can be applied in households where handicapped or elderly people live. A properly designed user interface along with ambient assisted living devices, such as Amazon's Alexa, which support voice commands,  will allow these people to modify easily the comfort parameters of their environment and the energy consumption will remain optimized.

\subsection{Advantages of Indirect Control IEMS}

\par The application of recommendation modules in IEMS is a very promising field of research. As it was mentioned in chapter 4, occupants' behavior is one  major influential  factor of inefficient energy consumption. Therefore, recommendation systems are useful tools to confront this issue. There is one significant advantage towards these systems. When the occupants follow recommendations for a long time, they acquire an eco-friendlier mindset. Thus, people trained from a home recommendation system will also apply the same aware behavior on every other aspect of their life, e.g. their workplace. Therefore, indirect control will eventually have wider impact.

\par A major difference from direct control systems is the lack of actuation infrastructure, which provides three important advantages. Firstly, these systems are cheaper and easier to buy. Deducting the cost of smart devices and appliances from the total cost of an IEMS, these systems become more affordable. Secondly, a system with less electrical parts connected to the internet, is less vulnerable to cyberattacks. Even if a false data injection attack is performed on them, the worst case scenario is a system that produces irrational recommendations. Moreover, these systems can also run using a local area network or a Bluetooth area network, which are more secure. Thirdly, the architecture of indirect control systems allows non-controllable appliances to take part into the optimization procedures without any modification on their hardware.

\par Last,  an additional feature of these systems is that it is possible to extend their design so that they can help consumers acquire appliances, based on input data from existing appliances in a household, power consumption of available models on the market, their price, manufacturer and other specifications (\cite{23}).

\subsection{Open Issues}

\par There has been great progress in the development of energy management systems, both towards direct and indirect control. However, various issues and limitations hinder the wide installation of such systems in commercial, residential and educational buildings. We will discuss the most important problematic aspects of each class.

\subsubsection{Security}

\par Direct control IEMS convert conventional buildings into smart ones. Therefore, they can be part of the demand-side of a smart grid, one of the most complex cyber-physical systems. According to  \cite{61}, every infrastructure based on cyber-physical systems is vulnerable to various types of attacks.

\par False Data Injection (FDI) is the most frequent type of attack. \cite{80}  proved that these attacks affect electricity bills and load consumption drastically, proposing  afterwards a resilient scheduling algorithm to overcome these effects.
Furthermore, \cite{81} showed that small fluctuations in energy demand from FDI attacks significantly increased the unit price and provided financial benefits to the attacker. Because of the nature of the demand response scheme, these situations lead to   inefficient energy consumptions by the system, increasing the energy footprint.

\par Additionally, another type of attack is the Denial of Service (DoS) attack. \cite{82} revealed a vulnerability of metering infrastructure using "Puppet Attack", a DoS attack that exhausts the communication bandwidth, proposing also detection and prevention mechanisms. Moreover,  \cite{83} demonstrated that some DoS attacks can cause disruption in the whole smart grid. However, they proposed an algorithmic solution to isolate the attacked nodes to continue data transmission.

\par  Other types of attacks are, Control Signal Adulteration (\cite{84}), which affects the automatic generation controller that regulates the frequency and power exchange between controlled areas and Information Leakage (\cite{85}).

\par  \cite{61} provided a classification of the attacks based on the part of the grid they occur. 

\begin{itemize}
    \item \textbf{Generation Systems attacks:} Attacks against power generation and power lines of the smart grid damaging the balance between generation and supply.
    
    \item \textbf{Transmission Systems attacks:} Attacks aiming to damage and interrupt the delivery of the generated energy through power stations and lines. These attacks can be classified as a) Interdiction attacks and b) Complex network attacks. (\cite{61})
    
    \item \textbf{End-user attacks:} Attacks on IoT devices and appliances at the end-user side, i.e. smart devices, appliances and electrical actuators. These attacks are serious, according to \cite{86}, because in smart metering and monitoring devices there are stored private information about users, such as user's activities, consumption and idle time and when user's location is empty or not. 
    
    \item \textbf{Electricity Market attacks:} Attacks that exploit vulnerabilities in the transmission management that affect the price of the electricity. That way, they make illegal profits and cause congestion to the power lines.
\end{itemize}

\subsection{Cold Start Problem \& Data Sparsity}

\par Cold Start Problem (CSP) refers to the lack of initial data  on a recommendation system. This issue occurs for various reasons, mostly in collaborative filtering models. Content-based and knowledge-based systems tend to be more robust (\cite{91}).

\par According to \cite{92}, there are three cases that cause CSP. First, the  \textit{new community} problem refers to the lack of ratings from the recommendation database. This usually occurs when there are not enough users to rate or vote for the proposed advice, therefore the precision of the recommendation can't be calculated. Second, the \textit{new item} problem. The new item problem occurs when new actions  or recommendations are imported in the database of the system. These recommendations contain no rating, therefore it is rare for them to be chosen. However, if a recommendation remains unnoticed for a long time, it acts like it doesn't exist at all, even though it could be useful. Finally, the greatest CS problem is the \textit{new user} problem. Someone who uses the system for the first time has zero votes on contained recommendation. The filtering methods of the system have no prior information about the new user and no history of ratings to calculate a "neighborhood" of appropriate recommendations (\cite{87}). Therefore, the performance of the system is negatively affected because it cannot produce meaningful recommendations (\cite{88}).

\par Some recommendation systems are based in collaborative filtering (CF), which  can be viewed as a classification and regression generalization (\cite{91}) and it is the most mature and commonly implemented technique (\cite{93}). This technique is making predictions about users' preferences based on data collected from users with similar profiles.  Ratings between users and items are stored in a matrix which is sparse, sometimes up to 99\% (\cite{94}). This problem is known as \textit{Data Sparcity} in CF recommendation systems.

\par Sparsity exists  when there is lack of knowledge about new users who start using the system and also, because they ignore the evaluation process after a recommendation. Moreover, new users rarely report their feedback on received suggestions. In CF systems, data sparsity is what causes the cold start problem ( \cite{94}).

\par To tackle these problems, some approaches have been proposed in the literature.  \cite{23} proposed that some explicitly stated preferences by the newcomers can be used as metrics to include them in some preexisting cluster of older users. \cite{89} proposed a model consisting of classification algorithms and similarity techniques that retrieved optimized recommendations. Furthermore, \cite{90} mentioned that promoting new items into new users is not very effective, but promoting new items into less active users showed some performance improvement. \cite{93} suggested an algorithm based on Sequence and Set Similarity Measure that utilized Singular Value Decomposition removes sparsity from the user-item-ratings matrix and  \cite{95} proposed two methods to overcome sparsity using linked open data from the  "DBpedia" knowledge base to create a recommendation system using  Matrix Factorization.

\subsection{Future Research}

\par This subsection discusses some insights for research to improve  direct and indirect control IEMSs in the  future.

\par As was mentioned earlier, privacy is a key aspect of IEMS. Protecting sensitive users' information is essential, therefore research is needed to improve their resilience to cyberattacks by developing frameworks with less vulnerabilities. Until now, due to different standards and communication stacks involved in the IoT technologies, traditional measure against cyberattacks are not always applicable (\cite{104}). Threats like systems' failure, smart meters' data corruption, infection by malware, spoofing of usernames and addresses and unauthorized access, show the necessity for research towards threat and risk modelling, IoT forensics, intrusion detection and prevention techniques (\cite{105}, \cite{106}).

\par  Furthermore, research is needed towards scalability of direct control IEMS. Expanding intelligent management of energy in the context of smart cities, indicates that smart systems should have the ability to be extended with various heterogeneous devices. Therefore, flexible layered architectures are necessary (\cite{107}) because non-scalable IEMS will be unreliable in the future (\cite{4}).

\par  In terms of indirect control systems, explainability of recommendations remains an issue that needs further study. The lack of explainable recommendations lead users to ignore advice, reducing the effectiveness of a system and the trustworthiness of it (\cite{110}).  Each recommendation should be accompanied by answers to the questions "why to perform an action" and "what the benefits are" (\cite{23}). In (EM)$^3$ system, (\cite{39,109}), each recommendation had a reasoning and a persuasion feature, which resulted in 20\% increase in the acceptance ratio.  Moreover, \cite{108} presented that it is more effective to provide justifications on  why an action should be performed rather than why not.

\par Gamified frameworks for indirect control of energy consumptions seem like  effective approaches. They are more engaging than conventional ones when they aim to change certain behaviors (\cite{113}). These approaches are preferred for school buildings, in order to improve energy awareness in students. An example by \cite{112}, was implemented  through the \cite{111} project, where  IoT lab kits where used along with a serious game resulting in acceptance of energy aware behaviors.

\begin{sidewaystable}
\sidewaystablefn%
\begin{center}
\begin{minipage}{\textheight}
\caption{Advantages of each IEMS class}\label{tab3}
\begin{tabular*}{\textheight}{@{\extracolsep{\fill}}p{8cm}p{8cm}@{\extracolsep{\fill}}}
\toprule%
Direct Control IEMS & Indirect Control IEMS    \\
\midrule
Automated procedures leave occupants' routine intact &  Behavioral changes leading to more eco-friendly mindsets have wider and long-term impact on the environment \vspace{0.5cm} \\

Provide plan and scheduling abilities to set long-term goals &  Human-in-the-loop applications offer cheaper implementations   \vspace{0.5cm} \\

Incorporate techniques to protect occupants in case of an emergency situation & Non-controllable appliances can take part in energy consumption management \vspace{0.5cm}\\

Accessibility features allow  installation in households with elderly or handicapped occupants & Extended designs can provide suggestions to replace appliances and devices based on users' profiles. \vspace{0.5cm}\\

\botrule
\end{tabular*}
\end{minipage}
\end{center}
\end{sidewaystable}

%% file: 7_conclusion.tex
\section{Conclusions}
\label{sec:conclusion}

\par This paper has reviewed  state-of-the-art  approaches of Intelligent Energy Management Systems. Within the area of energy efficiency, IEMS are considered as a way to confront climate change. These systems follow a similar architecture consisting of four components: Sensors, Actuators, Processing Engine and a User Interface. 

\par There are two types of sensing infrastructures in the literature,  custom-made and  commercial. Researchers choose their preferred type based on the goals and the scale of the application. In large-scale projects, commercial sensors provided convenience and sometimes a unified communication protocol, whereas custom made sensors were preferred for small-scale projects because they could embed more components.

\par Moreover, this review proposed a novel classification, based on the type of actuation. IEMS can be divided into direct and indirect control systems, depending on who is performing the  actions to optimize energy consumption. IEMS with incorporated automation modules are controlling the consumption directly, whereas IEMS aimed at behavioral changes suggest actions and allow the users to decide about actions. Direct control provides convenience  through automations and also safety in case of emergency situations. However, improving energy awareness through indirect control can bring about changes in larger scale.

\par Nevertheless, all of these systems have weak points and vulnerabilities. Systems with automations are mostly vulnerable against  cyberattacks. False Data Injection attacks in such systems, can cause an increase of consumed energy. Systems with recommendations  suffer from the Cold Start Problem which occurs when new users begin to use the system and when new actions are imported. These problems must be addressed to ensure the effectiveness of these applications.

\par To conclude, IEMS are going through a constant evolution. Direct control approaches seem like a better option for commercial buildings, where a large number of people  is present. In that scale, recommendation systems are not very promising. On the contrary, indirect control seems an appropriate choice for educational buildings because eventually, they will increase  the awareness of students and will provide  long term advantages. Finally, for residential environments, systems with automations are currently more advanced, however, the installation of a complete smart home is still very expensive and unaffordable for the majority of households. Therefore, we suggest that more indirect control applications must be developed for domestic environments in the future.